%% file: main.tex
\documentclass{Interspeech2024}

\interspeechcameraready

\usepackage{multirow,booktabs}
\usepackage{subfigure}

\title{MaskSR: Masked Language Model for Full-band Speech Restoration}

\name[affiliation={1 \star}]{Xu}{Li}
\name[affiliation={1,2 \star}]{Qirui}{Wang}
\name[affiliation={1 \star}]{Xiaoyu}{Liu}

\address{
$^1$Dolby Laboratories\\
$^2$School of Information Science and Engineering, Southeast University, China
}


\newcommand\blfootnote[1]{%
  \begingroup
  \renewcommand\thefootnote{}\footnote{#1}%
  \addtocounter{footnote}{-1}%
  \endgroup
}

\begin{document}

\maketitle
\ninept

\begin{abstract}
Speech restoration aims at restoring high quality speech in the presence of a diverse set of distortions. Although several deep learning paradigms have been studied for this task, the power of the recently emerging language models has not been fully explored. In this paper, we propose MaskSR, a masked language model capable of restoring full-band 44.1\,kHz speech jointly considering noise, reverb, clipping, and low bandwidth. MaskSR works with discrete acoustic tokens extracted using a pre-trained neural codec. During training, MaskSR is optimized to predict randomly masked tokens extracted from the high quality target speech, conditioned on the corrupted speech with various distortions. During inference, MaskSR reconstructs the target speech tokens with efficient iterative sampling. Extensive experiments show that MaskSR obtains competitive results on both the full-band speech restoration task and also on sub-tasks compared with a wide range of models.

\end{abstract}

\vspace{-1ex}
\section{Introduction}
\input{introduction}

\section{Method}
\input{method}

\vspace{-1ex}
\section{Experimental Setup}
\input{experiment_setup}

\section{Results}
\input{results}

\vspace{-1ex}
\section{Conclusion}
\input{conclusion}

\bibliographystyle{IEEEtran}
\bibliography{mybib}

\appendix
\input{appendix}

\end{document}

%% file: introduction.tex
\label{sec:introduction}

Speech restoration aims at restoring high quality speech from a corrupted input signal considering a diverse set of distortions~\cite{liu2022voicefixer, liu2023two, liu2024rad, chen2023gesper, kandpal2022music, serra2022universal, koizumi2023miipher}. Compared with conventional denoising and dereverberation, the diverse nature of the distortions (not just the quantity) makes this task much more challenging. Regression models succeed in removing noise and reverb~\cite{zhao2022frcrn, hao2021fullsubnet, li2021simultaneous}, but they cannot address tasks that are generative in nature, such as bandwidth extension, packet loss concealment, etc. To ease the task, a two-stage paradigm that employs separately trained models is widely adopted, in which one suppresses noise, and another one generates missing speech~\cite{liu2022voicefixer, liu2023two, liu2024rad, chen2023gesper, kandpal2022music}. Another variant jointly trains the two stages~\cite{serra2022universal}, but the success of the speech generation stage heavily relies on the previous stage that employs auxiliary losses to suppress the distortions. Thus, the power of generative models as a unified framework that addresses the considered distortions all at once has not been fully explored.\blfootnote{$^\star$ Equal contribution. Work done during Qirui's internship.}

Recently, language models (LMs) have gained popularity in audio and image synthesis due to their scalability, ease of training, and unification of different modalities as discrete tokens~\cite{kreuk2022audiogen, copet2024simple, deshmukh2024pengi, wang2023neural, chang2023muse}. Several works also show that LMs can translate noisy speech tokens to clean tokens end-to-end, providing an elegant framework~\cite{yang2023uniaudio, wang2023speechx, wang2023selm}, but only limited to denoising. In addition, to our knowledge, previous speech denoising LMs work with limited sampling rates up to 24\,kHz~\cite{wang2023speechx}. Therefore, the capability of LMs remains unknown for full-band speech restoration in the presence of a diverse set of distortions, all considered under a single generative framework. 

In this work, we propose MaskSR, a full-band 44.1\,kHz\footnote{Both 44.1 and 48\,kHz have been termed as full-band in previous research~\cite{zhao2022frcrn, liu22y_interspeech}. For brevity, we do not distinguish the two in this work.} speech restoration system that performs denoising, dereverberation, declipping, and bandwidth extension holistically. As shown in Figure~\ref{fig:LM_system}, MaskSR consists of a (frozen) pre-trained neural audio codec to (de)tokenize the high quality target speech, a speech encoder to encode a corrupted speech signal, and an LM conditioned on the encoded corrupted speech to predict the masked acoustic tokens of the target speech. During inference, MaskSR predicts the target speech tokens with efficient iterative sampling. MaskSR obtains strong results on both the contributed full-band speech restoration task evaluated with a blend of the studied distortions, and also on individual tasks compared with a wide range of models.

%% file: method.tex
\label{sec:Method}

\begin{figure*}[t]
  \centering
  \vspace{-2ex}
  \includegraphics[width=\textwidth]{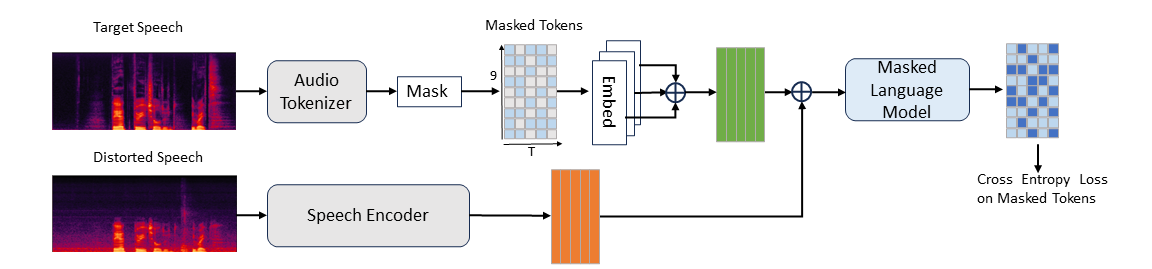}
  \vspace{-3ex}
  \caption{Training workflow of MaskSR.}
  \label{fig:LM_system}
  \vspace{-3ex}
\end{figure*}

\subsection{Neural Audio Tokenizer}
\label{sec:tokenizer}

We use the (frozen) pre-trained Descript Audio Codec (DAC) as our speech (de)tokenizer~\cite{kumar2024high}. DAC is a state-of-the-art auto-encoder. In the training stage of MaskSR, the DAC encoder projects a 44.1\,kHz high quality target speech signal to $T$ frames in a latent space with a reduced sampling rate of $\sim$86\,Hz. Each frame is then tokenized by 9 residual vector quantizers (RVQs)~\cite{zeghidour2021soundstream}. Each RVQ quantizes the error of the previous one with a codebook size 1024. Thus, a waveform becomes a $9\times{T}$ codegram. During inference, the DAC decoder detokenizes a codegram predicted by the LM to a waveform. DAC is pre-trained to accurately reconstruct the unquantized waveform. Note that although MaskSR can be divided into two stages: DAC and the rest, it's fundamentally different than the previous two-stage systems that split the removal of various distortions across cascaded models. Here, DAC only creates a compact discrete space that encodes enough acoustic details of the full-band speech to be restored, and it's the rest of the system that performs the restoration jointly considering all the distortions.   

\subsection{Speech Encoder}
\label{sec:encoder}

The speech encoder first computes the power-law compressed magnitude STFT spectrogram $X^{0.3}$ given a corrupted speech signal $x$ using a window length and hop length of 2048 and 512 samples, respectively. The hop length is consistent with that of the DAC encoder to align the STFT and DAC frames along time. Next, a multi-layer perceptron (MLP) followed by a stack of self-attention transformer blocks map the STFT features to $d$ dimensional embeddings compatible with the DAC space, so that the speech encoder and the LM could be jointly optimized. 

\subsection{Masked Language Model}
\label{sec:masklm}

Inspired by~\cite{serra2023mono, garcia2023vampnet}, we extend the Masked Generative Image Transformer (MaskGIT)~\cite{chang2022maskgit} originated from modeling 1-D sequences to 2-D codegrams. During training, after obtaining a $9\times{T}$ codegram from a full-band target speech, we randomly mask a subset of tokens by replacing them with a special mask token. Then, we use 9 learnable embedding tables each with 1025 entries (including the mask token) to embed the 9 codebooks, respectively, each resulting in a $T\times{d}$ tensor. Inspired by~\cite{serra2023mono}, we aggregate the codebooks by summing the 9 tensors, and also sum the resulting embeddings with the speech encoder representation of the corrupted speech. The summation of the codebook embeddings keeps the sequence length unchanged despite of multiple codebooks, thus minimizes the system complexity. The LM uses a stack of self-attention transformer blocks to model the aggregated sequences. Due to masking, the LM is free to learn from all the positions in the codegram, as opposed to autoregressive LMs~\cite{copet2024simple, yang2023uniaudio}. Finally, an output layer consisting of 9 1024-dim softmax classifiers computes the logit scores of the tokens from the 9 codebooks. The LM and the speech encoder are jointly optimized with a cross entropy loss only applied to the masked positions.

During inference, starting from a fully masked codegram, we conduct iterative sampling as in~\cite{chang2022maskgit} to gradually generate the target speech tokens. In each iteration, the LM predicts 1024-dim token probability distributions in all the masked positions in parallel, given the tokens generated from previous iterations. We sample a token in each masked position from the predicted distribution, and re-mask a subset of the sampled tokens with low logit scores. The percentage of re-masking is controlled by a cosine schedule. We add Gaussian noise to the logit scores before ranking them to increase diversity, and the variance linearly decreases from 4 to 0 throughout inference. We perform 40 iterations until the codegram is fully reconstructed.

In addition, we use classifier-free guidance formulated for LMs~\cite{chang2023muse, sheffer2023hear} on the speech encoder representation of the corrupted speech to prevent the generated speech from deviating too far from the original speaker voice. During training, we randomly replace the speech encoder output 10\% of the time with a learnable embedding repeated $T$ times, and during inference, the logit scores $l_g$ of the tokens are computed as:

\vspace{-1ex}
\begin{equation*}
    l_g = (1+w)l_c - wl_u
\end{equation*}

\noindent where $l_c$ and $l_u$ are the conditional and unconditional logits, respectively. A larger guidance $w\ge0$ improves the speaker identity preservation at the cost of slightly more residual noise.

\subsection{Other Codebook Modeling Strategies}
\label{sec:other_lm}

Since efficiently and effectively modeling multiple codebooks is crucial to the system performance, we compare the parallel strategy in MaskSR with another 2 representative variants.

\textbf{SoundStorm}~\cite{borsos2023soundstorm} exploits the hierarchical nature of the RVQ, noting that the low level codebooks help predicting the higher level ones. For each training sample, only the tokens from a randomly selected codebook are randomly masked and predicted in the MaskGIT fashion, whereas all the lower codebooks are assumed available, and all the higher level codebooks are fully masked (but not predicted). During inference, the codebooks are reconstructed hierarchically, with the first one based on the MaskGIT iterative sampling, and the rest simply taking the tokens with the highest probabilities. SoundStorm only requires running the LM 8 more times (with 9 codebooks) compared with MaskSR, thus only modestly increases inference time. But we observe that the first codebook, which encodes the most salient speech patterns, is not well modeled (Sec.~\ref{sec:full_restore}).

\textbf{UniAudio}~\cite{yang2023uniaudio} is a recent LM that runs autoregressively along both the time and codebook dimensions. Thus, it is much slower than MaskSR and SoundStorm during inference. Also, the causal transformers in UniAudio only have access to the past tokens which may hurt the modeling capability whereas masked LMs do not have such a limitation.

%% file: experiment_setup.tex
\label{sec:setup}

\begin{table*}[t]
    \renewcommand\thetable{1}
    \caption{Full-band 44.1\,kHz speech restoration results on the SR and ALL-GSR test sets}
    \centering
    \resizebox{\textwidth}{!}{
    \begin{tabular}{@{}cccccccccccccc@{}}
    \toprule
    \multirow{3}{*}{System} & \multirow{3}{*}{Model size} &
      \multicolumn{6}{c}{SR clean test set for bandwidth extension} &
      \multicolumn{6}{c}{ALL-GSR test set with all 4 studied distortions} \\
    \cmidrule(lr){3-8} \cmidrule(lr){9-14}
     & & \multicolumn{3}{c}{DNSMOS $\uparrow$} & \multirow{2}{*}{SESQA $\uparrow$} & \multirow{2}{*}{LSD $\downarrow$} &
     \multirow{2}{*}{Spk Sim $\uparrow$} &
     \multicolumn{3}{c}{DNSMOS $\uparrow$} & \multirow{2}{*}{SESQA $\uparrow$} & \multirow{2}{*}{LSD $\downarrow$} & \multirow{2}{*}{Spk Sim $\uparrow$} \\
    \cmidrule(lr){3-5} \cmidrule(lr){9-11}
     & & SIG & BAK & OVL & & & & SIG & BAK & OVL & & & \\
    \midrule
    Unprocessed & - & 3.413 & 4.025 & 3.107 & 2.577 & 2.889 & 0.808 & 2.961 & 2.857 & 2.393 & 2.598 & 2.014 & \textbf{0.901} \\
    Target-DAC & - & 3.472 & 4.044 & 3.174 & 3.488 & 0.837 & 0.933 & 3.455 & 3.981 & 3.143 & 3.533 & 0.827 & 0.931 \\ 
    \midrule
    NSNet2 & 2.8\,M & 2.947 & \textbf{4.077} & 2.584 & 2.933 & 2.868 & 0.741 & 3.001 & 3.983 & 2.749 & 3.010 & 2.545 & 0.867 \\
    VoiceFixer & 111\,M & 3.401 & 4.039 & 3.109 & 3.339 & 1.044 & 0.737 & 3.298 & 3.969 & 3.002 & 3.396 & \textbf{1.019} & 0.781 \\
    \midrule
    SoundStorm & 55\,M & 3.423 & 4.001 & 3.117 & 3.426 & 1.062 & 0.812 & 3.395 & 3.973 & 3.085 & 3.485 & 1.171 & 0.833 \\
    UniAudio & 55\,M & 3.415 & 4.022 & 3.110 & 3.447 & 1.036 & 0.792 & 3.403 & \textbf{4.026} & 3.117 & 3.538 & 1.363 & 0.815 \\
    \midrule
    MaskSR-S & 55\,M & \textbf{3.442} & 4.017 & 3.135 & 3.430 & 0.978 & 0.822 & 3.430 & 3.982 & 3.123 & \textbf{3.541} & 1.201 & 0.845 \\
    MaskSR-M & 145\,M & 3.440 & 4.021 & \textbf{3.136} & \textbf{3.467} & \textbf{0.959} & \textbf{0.832} & \textbf{3.445} & 3.971 & \textbf{3.128} & 3.531 & 1.191 & 0.853 \\
    \bottomrule
    \end{tabular}
    }
    \label{tab:fullband_results}
    \vspace{-3ex}
\end{table*}

\subsection{Datasets}
\label{sec:data}

\textbf{Training set}\quad
We use $\sim$800 hours of publicly available clean speech including the `read speech' and VCTK~\cite{veaux2017cstr} subsets provided by the 2022 DNS Challenge~\cite{dubey2022icassp}, and also the AISHELL-1 dataset~\cite{bu2017aishell}. We use 181 hours of noise and 60\,k room impulse responses (RIRs) also from~\cite{dubey2022icassp}. All speech, noise, and RIRs are recorded with 48\,kHz or 44.1\,kHz sampling rates, and we downsample the data from 48\,kHz to 44.1\,kHz to be compatible with DAC. We consider 4 types of distortions as in~\cite{liu2022voicefixer}: noise, reverb, clipping, low bandwidth, and create 44.1\,kHz corrupted speech on the fly using the open-source pipeline in~\cite{liu2022voicefixer}, resulting in distorted samples with an SNR in $[-5, 20]$\,dB, clipped between $[0.1, 0.5]$, and a bandwidth from 1\,kHz to 22.05\,kHz. 

\vspace{1ex}
\noindent \textbf{Full-band test sets}\quad
We use the 44.1\,kHz open-source SR and ALL-GSR test sets used by~\cite{liu2022voicefixer} to evaluate full-band models. SR contains clean data with bandwidth between 1\,kHz and 8\,kHz, targeting at bandwidth extension only. ALL-GSR contains a blend of the 4 studied distortions. Overlapping speakers that also appear in VCTK are excluded from the training set.

\vspace{1ex}
\noindent \textbf{Wide-band test sets}\quad
To compare extensively with the majority of models that only perform denoising and/or dereverberation at 16\,kHz, we consider the 2020 DNS Challenge~\cite{reddy2020interspeech} test sets, including the synthetic data with and without reverb, and the real recordings. To run MaskSR, we upsample the input speech to 44.1\,kHz, and downsample the output back to 16\,kHz.

\subsection{Implementation Details}
\label{sec:implement}

We use the pre-trained DAC released in~\cite{kumar2024high} as our speech tokenizer. A small version MaskSR-S uses an embedding dimension $d=512$, sinusoidal positional encoding, and there are 6 and 8 transformer blocks in the speech encoder and the LM, respectively, each with 16 attention heads, an MLP with a hidden dimension $4d$, and pre-norm. A medium size MaskSR-M uses $d=768$ and 12 transformer blocks in the LM. SoundStorm and UniAudio share the same overall system (Figure~\ref{fig:LM_system}) and model size as MaskSR-S to fairly compare codebook modeling methods. We use the official UniAudio implementation in~\cite{yang2023uniaudio}. All models are trained on 3\,sec speech segments for 800\,k steps on 4 A100 GPUs with a learning rate of 0.0001 using the Adam optimizer. We use a batch size 256 for \mbox{MaskSR-M} and 64 for other models. During inference, we decode each non-overlapping 3\,sec window with 40 and 48 iterations for MaskSR and SoundStorm, respectively, and 2331 iterations ($9\times{259}$) for UniAudio.

\subsection{Baseline Models}
\label{sec:baseline}

\noindent \textbf{Full-band models}\quad
In addition to the 3 LMs, we consider 2 full-band models: VoiceFixer~\cite{liu2022voicefixer} and NSNet2~\cite{braun2020data}. VoiceFixer is a strong 2-stage speech restoration model targeting at the same 4 distortions as in our work. NSNet2 is a regression-only model provided as the DNS Challenge baseline, performing denoising and dereverberation. We use the model checkpoints from~\cite{liu2022voicefixer, dubey2022icassp}. The released VoiceFixer was trained on a different dataset, but re-training VoiceFixer on our data did not obtain better results on the full-band test sets. Thus, we stick with the official checkpoint to report the results.

\vspace{1ex}
\noindent \textbf{Wide-band models}\quad
MaskSR is compared with a collection of models specializing in denoising on the 16\,kHz wide-band test sets. We obtain the released DEMUCS and FRCRN checkpoints from~\cite{defossez2020real, zhao2022frcrn} as strong regression candidates. We use the results reported in Wang et al.~\cite{wang2023selm} for SGMSE~\cite{richter2023speech}, StoRM~\cite{lemercier2023storm}, and SELM~\cite{wang2023selm}. The former two are diffusion models and the third is a recent speech enhancement LM. All the models are trained on datasets comparable to ours. In addition, DEMUCS also jointly performs dereverberation.

\vspace{1ex}
\noindent \textbf{Unprocessed and Target-DAC} refer to the corrupted input speech and DAC-processed target speech, respectively. Target-DAC is an upper bound for the LM-based models studied in this work that employ DAC as the tokenizer.

\subsection{Evaluation Metrics}
\label{sec:metric}

It's a known fact that standard metrics such as PESQ, SI-SNR cannot accurately assess generative models due to lack of waveform alignment~\cite{serra2022universal, jassim2021warp}. We rely on the following metrics instead, and resample the generated speech if necessary.

\vspace{1ex}
\noindent \textbf{DNSMOS and SESQA} are reference-free perceptual quality estimators~\cite{reddy2021dnsmos, serra2021sesqa} capable of evaluating generative models aiming to fix similar distortions~\cite{serra2022universal, wang2023selm, pascual2021adversarial}. SESQA works with 48\,kHz and DNSMOS works with 16\,kHz. We use the public DNSMOS~\cite{reddy2021dnsmos} and our in-house SESQA trained on the data and model configurations as described in~\cite{serra2021sesqa}.

\vspace{1ex}
\noindent \textbf{Log-Spectral Distance (LSD)}~\cite{erell1990estimation} is a common metric to measure bandwidth extension. LSD supports 44.1\,kHz. We use the public implementation in~\cite{liu2022voicefixer}.

\vspace{1ex}
\noindent \textbf{Speaker Similarity (Spk Sim)} is the speaker cosine similarity between the ground truth and the processed speech. We use the public WeSpeaker~\cite{wang2023wespeaker} to compute similarity at 16\,kHz.

\vspace{1ex}
\noindent \textbf{Subjective Listening (MOS)}\quad
We ask 14 expert listeners to rate the overall generated speech quality on a 1--5 scale, and report the mean opinion scores based on 40 samples from the full-band ALL-GSR test set that covers the 4 studied distortions and their combinations. Samples are available on our demo page\footnote{https://masksr.github.io/MaskSR/}.

%% file: results.tex
\label{sec:results}

\begin{table*}[t]
    \renewcommand\thetable{4}
    \vspace{-2ex}
    \caption{Wide-band 16\,kHz denoising/dereverberation results on the DNS Challenge test sets. The SGMSE, StoRM, and SELM results are reported in~\cite{wang2023selm}. On the `With Reverb' test set, only MaskSR and DEMUCS perform joint denoising and dereverberation while other models only perform denoising (see Sec.~\ref{sec:16k_results}).}
    \centering
    \resizebox{\textwidth}{!}{
    \begin{tabular}{@{}ccccccccccccc@{}}
    \toprule
    \multirow{3}{*}{System} & \multirow{3}{*}{Model type} &
      \multicolumn{4}{c}{With Reverb} &
      \multicolumn{4}{c}{Without Reverb} &
      \multicolumn{3}{c}{Real Recordings} \\
    \cmidrule(lr){3-6} \cmidrule(lr){7-10} \cmidrule(lr){11-13}
     & & \multicolumn{3}{c}{DNSMOS $\uparrow$} & \multirow{2}{*}{Spk Sim $\uparrow$} &
     \multicolumn{3}{c}{DNSMOS $\uparrow$} & \multirow{2}{*}{Spk Sim $\uparrow$} &
     \multicolumn{3}{c}{DNSMOS $\uparrow$} \\
    \cmidrule(lr){3-5} \cmidrule(lr){7-9} \cmidrule(lr){11-13}
     & & SIG & BAK & OVL & & SIG & BAK & OVL & &  SIG & BAK & OVL \\
    \midrule
    Unprocessed & - & 1.760 & 1.497 & 1.392 & \textbf{0.941} & 3.392 & 2.618 & 2.483 & 0.969 & 3.053 & 2.509 & 2.255 \\ 
    \midrule
    DEMUCS & Regression & 2.856 & 3.897 & 2.553 & 0.762 & 3.575 & \textbf{4.153} & \textbf{3.345} & 0.956 & 3.263 & \textbf{4.027} & 2.988 \\
    FRCRN & Regression & 2.934 & 2.924 & 2.279 & 0.935 & 3.578 & 4.133 & 3.335 & \textbf{0.970} & 3.370 & 3.977 & 3.037 \\
    \midrule
    SGMSE & Diffusion & 2.730 & 2.741 & 2.430 & - & 3.501 & 3.710 & 3.137 & - & 3.297 & 2.894 & 2.793 \\
    StoRM & Diffusion & 2.947 & 3.141 & 2.516 & - & 3.514 & 3.941 & 3.205 & - & 3.410 & 3.379 & 2.940 \\
    \midrule
    SELM & LM & 3.160 & 3.577 & 2.695 & - & 3.508 & 4.096 & 3.258 & - & \textbf{3.591} & 3.435 & 3.124 \\
    \midrule
    MaskSR-S & LM & 3.524 & 4.016 & 3.223 & 0.816 & 3.575 & 4.082 & 3.307 & 0.926 & 3.398 & 4.011 & 3.103 \\
    MaskSR-M & LM & \textbf{3.531} & \textbf{4.065} & \textbf{3.253} & 0.827 & \textbf{3.586} & 4.116 & 3.339 & 0.929 & 3.430 & 4.025 & \textbf{3.136} \\
    \bottomrule
    \end{tabular}
    }
    \label{tab:wideband_results}
    \vspace{-2ex}
\end{table*}

\subsection{Full-band 44.1 kHz Speech Restoration}
\label{sec:full_restore}

\begin{figure}[t]
  \centering
  \includegraphics[scale=0.4]{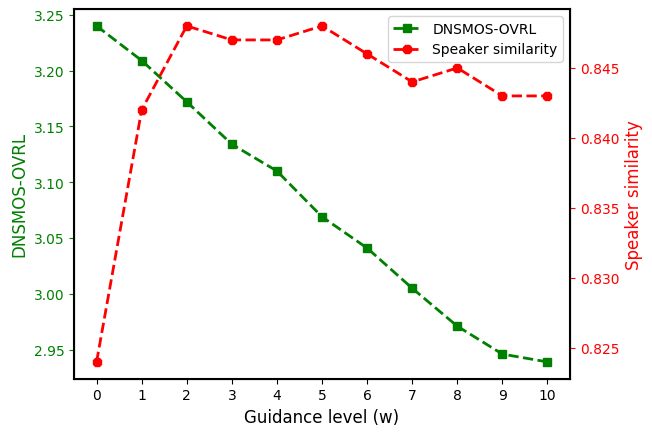}
  \vspace{-2ex}
  \caption{Effects of guidance on the overall DNSMOS (green) and speaker similarity scores (red).}
  \label{fig:guidance}
  \vspace{-4ex}
\end{figure}

First, we show the effects of the guidance level $w$ on a held-out dev set. Figure~\ref{fig:guidance} shows that a larger $w$ yields the peak speaker similarity at $w=2$ due to more alignment with the input speech, but DNSMOS decreases due to more residual noise. This shows the complementary nature of the two scores. We use $w=2$ to report all the results without tuning on the test sets.

In Table~\ref{tab:fullband_results}, on the SR clean test set, since DNSMOS is not sensitive to bandwidth extension, we mainly rely on the other 3 scores. We see that both MaskSR models achieve leading bandwidth extension performance. They are better than the two-stage VoiceFixer, and the regression-based NSNet2, which cannot perform this task. On the ALL-GSR test set with a blend of all the 4 studied distortions, both MaskSR also obtain competitive results, which outperform VoiceFixer in terms of all scores except for LSD. This indicates the strong capability of end-to-end speech restoration in the discrete space.

\begin{figure}[t]
  \centering
  \includegraphics[scale=0.38]{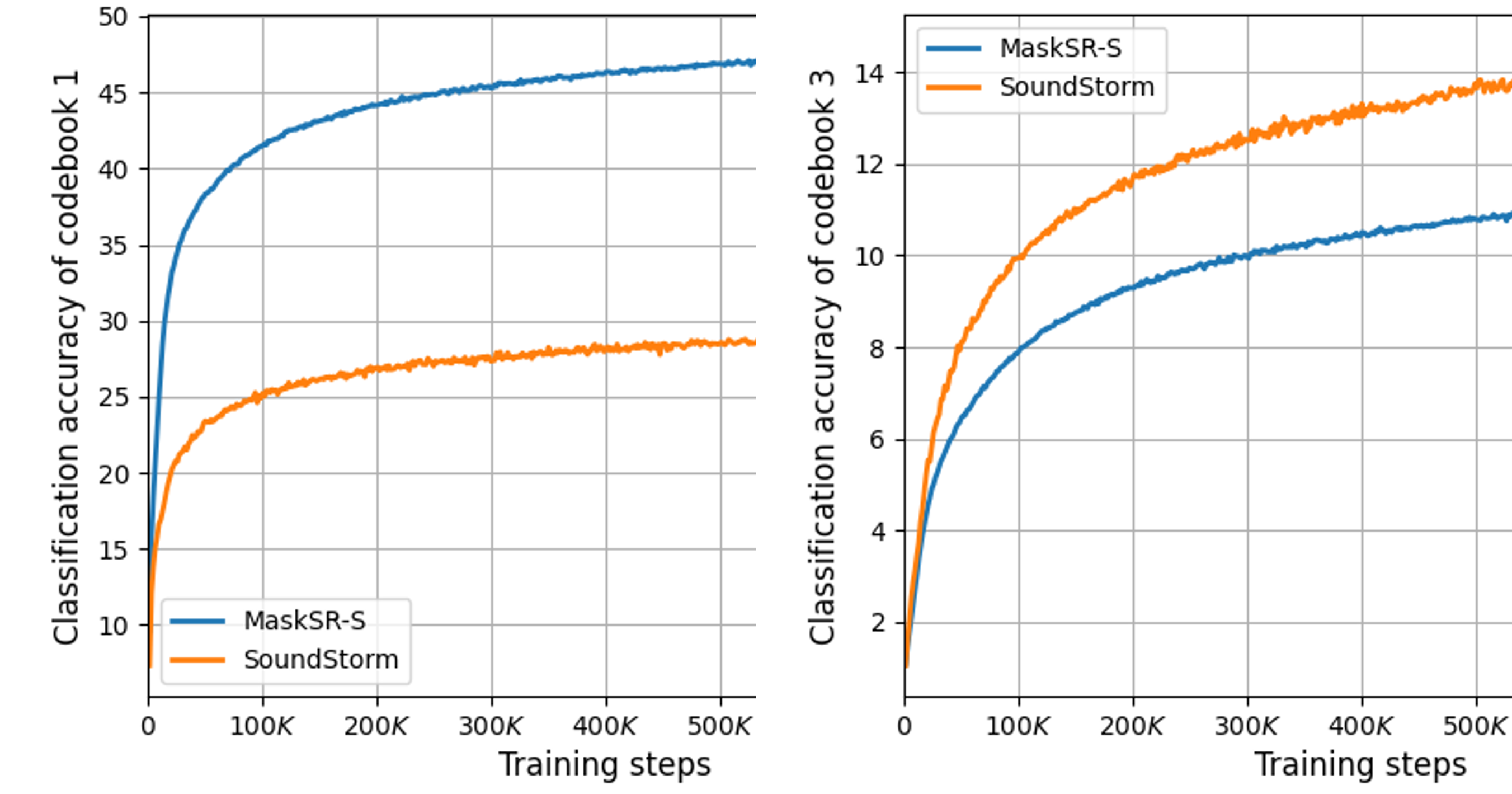}
  \vspace{-1ex}
  \caption{Token classification accuracy of codebook 1 (left) and 3 (right) from MaskSR-S (blue) and SoundStorm (orange). Other codebooks follow the trend of codebook 3.}
  \label{fig:acc}
  \vspace{-5ex}
\end{figure}

We notice that SoundStorm, in terms of all results except for the ALL-GSR LSD, does not outperform MaskSR-S. In Figure~\ref{fig:acc}, we compare the two models in terms of the token accuracy from codebook 1 and 3. Due to the hierarchical codebook modeling (Sec.~\ref{sec:other_lm}), SoundStorm yields modestly higher accuracy for codebook 2--9 (represented by codebook 3) than that of MaskSR-S, at the cost of significantly lower codebook 1 accuracy, the codebook that encodes the most salient speech pattern, such as speaker identity. This is because that no other codebooks are available when predicting codebook 1. The much lower codebook 1 accuracy may lead to the consistently worse speaker similarity scores in Table~\ref{tab:fullband_results}, whereas the modestly higher codebook 2--9 accuracy does not consistently translate to better LSD that reflects the generated high frequency details. Thus, the fully parallel codebook modeling in MaskSR provides overall better performance. In addition, we also measure the average runtime (over 20 runs) of the studied LMs on an A100 GPU. Table~\ref{tab:runtime} shows that MaskSR-S is slightly faster than SoundStorm, and significantly faster than the autoregressive UniAudio, which also does not yield better quality.   

\begin{table}[t]
  \renewcommand\thetable{2}
  \caption{Average runtime (sec) of different language models}
  \label{tab:runtime}
  \centering
  \begin{tabular}{ c c  c c c }
    \toprule
    \multirow{2}*{System} &
    \multicolumn{4}{c}{Sequence length (sec)} \\
    \cmidrule{2-5}
        &  4    & 8    &  12   &    16         \\
    \midrule
    UniAudio & 44.86   & 66.08    &  87.54 & 112.02   \\
    SoundStorm & 3.05    &  4.25   & 5.55 & 6.85 \\  
    MaskSR-S & \textbf{2.72}   &  \textbf{4.10}   & \textbf{4.92} & \textbf{5.22} \\     
    \bottomrule
  \end{tabular}
\end{table}

\begin{table}[t]
  \renewcommand\thetable{3}
  \caption{Subjective MOS scores with 95\% confidence intervals}
  \vspace{-1ex}
  \label{tab:listen_result}
  \centering
  \resizebox{1.0\columnwidth}{!}{
  \begin{tabular}{ c  c c c c }
    \toprule
    Unprocessed & Target & NSNet2 & VoiceFixer & MaskSR-M \\
    \midrule
    {1.96 $\pm$ 0.08} & {4.66 $\pm$ 0.05} & {2.50 $\pm$ 0.09} & {3.53 $\pm$ 0.09} & \textbf{4.36 $\pm$ 0.07} \\
    \bottomrule
  \end{tabular}
  }
  \vspace{-4ex}
\end{table}

Table~\ref{tab:listen_result} reports the subjective listening results based on the 40 samples from the ALL-GSR test set. The MOS reflects the overall speech quality. MaskSR-M significantly outperforms other systems, showing superior capability to restore high quality full-band speech from diverse distortions.  

\vspace{-1ex}
\subsection{Wide-band 16 kHz Denoising and Dereverberation}
\label{sec:16k_results}

In Table~\ref{tab:wideband_results}, on the `With Reverb' test set, since only MaskSR and DEMUCS~\cite{defossez2020real} perform joint noise and reverb suppression while other models only suppress noise, this partially contributes to the higher DNSMOS for MaskSR and DEMUCS. But comparing only these two, MaskSR still outperforms DEMUCS by a large margin. Meanwhile, since the ground truth contains reverb, but the outputs of MaskSR and DEMUCS do not, that leads to lower speaker similarity scores relative to other denoising-only models. On the other two test sets without noticeable reverb, despite the fact that MaskSR is trained to address a diverse set of distortions, it still achieves competitive denoising results compared to various specialized models.

%% file: conclusion.tex
\label{sec:conclusion}

In this work, we proposed a full-band speech restoration system that addressed a diverse set of distortions holistically based on masked LMs. The system showed promising results on both the full-band speech restoration task evaluated with a blend of the studied distortions, and also on individual tasks. We also shed insights into the effects of codebook modeling on the studied task, and improved speaker identity preservation using classifier-free guidance. Our future work includes further improving the generated speech quality and intelligibility by exploring semantic tokens~\cite{wang2023selm}.

%% file: appendix.tex
\begin{table*}
    \renewcommand\thetable{5}
    \captionsetup{width=0.9\textwidth}
    \caption{ALL-GSR full-band speech restoration results using different input features to encode the corrupted speech}
    \centering
    \resizebox{0.9\textwidth}{!}{
    \begin{tabular}{cccccccc}
    \toprule
    \multirow{2}{*}{Input feature} & \multirow{2}{*}{Model size} &
    \multicolumn{3}{c}{DNSMOS $\uparrow$} & \multirow{2}{*}{SESQA $\uparrow$} & \multirow{2}{*}{LSD $\downarrow$} &
     \multirow{2}{*}{Spk Sim $\uparrow$} \\
    \cmidrule(lr){3-5}
     & & SIG & BAK & OVL & & &\\
    \midrule
    DAC &  54\,M & 3.179 & 3.915 & 2.871 & 3.440 & 1.286 & 0.803 \\
    Waveform & 59\,M & 3.377 & 3.973 & 3.068 & \textbf{3.573} & 1.276 & 0.837 \\
    STFT (MaskSR) & 55\,M & \textbf{3.430} & \textbf{3.982} & \textbf{3.123} & 3.541 & \textbf{1.201} & \textbf{0.845} \\
    \bottomrule
    \end{tabular}
    }
    \label{tab:encoder_in_domain}
\end{table*}

\section{Input Speech Representation}
\label{sec:enc_domain}

Previous speech denoising LMs~\cite{yang2023uniaudio, wang2023speechx} encode the input corrupted speech as discrete tokens extracted from a pre-trained audio tokenizer. Since these models target at translating various input modalities (such as text and audio) to a target audio under a unified multi-task framework, it's convenient to encode all types of input signals as discrete tokens. However, for the dedicated speech restoration task, it's unclear whether such representation of the corrupted speech is optimal or not. The lossy compression caused by the neural audio tokenizer could hurt the integrity of the crucial information in the input speech, such as low level speaker characteristics, high frequency details, etc. Therefore, it might be beneficial to use lossless transformations to encode the input speech. To study the effects of the input features, we consider 3 options.

\vspace{1ex}
\noindent \textbf{DAC} We extract the $9\times{T}$ codegram from the input corrupted speech using the pre-trained DAC tokenizer. Then, following the same method to embed the target speech codegram (Sec.~\ref{sec:masklm}), we obtain the summation of the codebook embeddings from the 9 learnable embedding tables. Since the corrupted and the target speech share the same DAC space, we directly sum their embeddings without using a transformer-based speech encoder to further transform the corrupted speech. For a fair comparison, we use 14 transformer blocks in the LM with approximately the same total model capacity as other variants which do employ a speech encoder.

\vspace{1ex}
\noindent \textbf{STFT} This is the adopted method in MaskSR. As detailed in Sec.~\ref{sec:encoder}, we compute the power-law compressed magnitude STFT spectrogram given a corrupted speech signal using a window length and hop length of 2048 and 512 samples, respectively. Next, an MLP followed by a stack of 6 self-attention transformer blocks map the STFT features to 512-dim embeddings, which are summed with those of the masked target speech. An LM consisting of 8 transformer blocks is used to predict the masked target speech tokens.

\vspace{1ex}
\noindent \textbf{Waveform} We replace the STFT by a learnable 2048-dim \mbox{1-D} convolution followed by a ReLU activation function and a layer normalization. To be consistent with STFT, the convolution also uses a kernel length and stride of 2048 and 512 samples, respectively. After the layer normalization, the 2048-dim latent features of the corrupted speech signal are projected down to 512-dim embeddings by a fully connected layer followed by 6 self-attention transformer blocks. The LM also uses 8 transformer blocks as in the STFT model. Compared with DAC, both STFT and waveform are lossless representations of the input speech. 

We train the 3 systems on the dataset described in Sec.~\ref{sec:data}, and evaluate them on the full-band ALL-GSR test set. We optimize the classifier-free guidance level $w$ for each of them during inference on a held-out dev set, and choose $w=0.3$ for the DAC model and $w=2$ for the other two variants.   

From Table~\ref{tab:encoder_in_domain}, it can be seen that there is a noticeable gap between the system that uses DAC to encode the input speech and the other two variants that employ raw features. Although the LM component in the DAC-based model is larger than those in the other two systems (14 vs.~8 transformer blocks), the quality of its generated speech is lagging. This shows that the discrete DAC token is not the optimal feature representation to enable high quality speech restoration. On the other hand, the STFT model performs better than the waveform model in terms of most metrics. Empirical listening finds that the advantage of the STFT model is larger on the dereverberation task as it generates more natural speech with less over-suppression of the target speech and better speaker voice preservation. Thus, these results provide the supporting evidence for adopting STFT in MaskSR.